\documentclass[aps,prd,preprint,nofootinbib,amssymb]{revtex4}

\usepackage{amsmath,setspace,subfigure,amsfonts,latexsym}
\usepackage{amssymb}
\usepackage{color}
\usepackage{epsfig}
\usepackage{color}
\definecolor{White}{rgb}{1,1,1}
\definecolor{Red}{rgb}{1,0.1,0}
\definecolor{LightYellow}{rgb}{1,1,.875}
\definecolor{SteelBlue}{rgb}{.273,.508,.703}
\definecolor{navy}{rgb}{0,0,.5}
\definecolor{LightCyan}{rgb}{.875,1,1}
\definecolor{DarkRed}{rgb}{.543,0,0}
\definecolor{HotPink}{rgb}{1,.41,.70}
\definecolor{ForestGreen}{rgb}{.13,.54,.13}
\definecolor{OliveDrab}{rgb}{.42,.55,.14}
\definecolor{MediumBlue}{rgb}{0,0,.80}
\definecolor{RoyalBlue}{rgb}{.25,.41,.88}
\definecolor{DeepSkyBlue}{rgb}{0,.746,1}
\definecolor{Brown}{rgb}{0.545,0.271,0.074}

\def\bea{\begin{eqnarray}}
\def\eea{\end{eqnarray}}
\def\bec{\begin{center}}
\def\ec{\end{center}}

\def\gtr{\gtrsim}
\def\beq{\begin{equation}}
\def\eeq{\end{equation}}

\def\gtr{\gtrsim}

\newcommand\lsim{\mathrel{\rlap{\lower4pt\hbox{\hskip1pt$\sim$}}
    \raise1pt\hbox{$<$}}}
\newcommand\gsim{\mathrel{\rlap{\lower4pt\hbox{\hskip1pt$\sim$}}
    \raise1pt\hbox{$>$}}}
\def\bea{\begin{eqnarray}}
\def\eea{\end{eqnarray}}
\def\ba{\begin{array}}
\def\ea{\end{array}}
\def\bc{\begin{center}}
\def\ec{\end{center}}

\begin{document}

\title{Cosmology of the DFSZ axino}

\author{Kyu Jung Bae$^a$\footnote{kyujung.bae@kaist.ac.kr}, Eung Jin Chun$^b$\footnote{ejchun@kias.re.kr} and Sang Hui Im$^a$\footnote{shim@muon.kaist.ac.kr}}

\address{$^a$Department of Physics, KAIST, Daejeon 305-701, Korea\\
$^b$Korea Institute for Advanced Study, Seoul 130-722, Korea}

\begin{abstract}
We study the cosmological impact of the supersymmetric DFSZ axion model.
Extending recent works, we first provide a comprehensive analysis
of thermal production of the DFSZ axino considering all the possible scattering,
decay and inverse decay processes depending on various mass parameters and the reheat temperature.
Although it is hard for the DFSZ axino to be in thermal equilibrium,
its coupling is still large enough to generate huge axino population which
can turn into overabundant neutralino density.
We examine the neutralino parameter space to identify the dark matter property depending on
the Peccei-Quinn scale. As the Peccei-Quinn scale becomes higher resulting in
longer axino lifetime,
the neutralino dark matter appears
in a lighter Higgsino-like LSP region or a more restricted Bino-like LSP region
allowing a resonant annihilation through a CP-odd Higgs boson to meet stronger
reannihilation.
\end{abstract}

\maketitle

\section{Introduction}

The most attractive solution to the strong CP problem would be to introduce the axion,
a pseudo-Goldstone boson of the Peccei-Quinn (PQ) symmetry which is supposed to be broken
at $v_{PQ} = 10^{9}-10^{12}$ GeV \cite{Kim08}.
The axion solution can be realized typically in two ways: (i) the KSVZ model assuming a heavy quark
at an intermediate scale $\lesssim v_{PQ}$ \cite{ksvz}; (ii) the DFSZ model extending the Higgs sector with PQ charged singlet fields \cite{dfsz}.  In the supersymmetric extension of the axion models, the presence of the axino $\tilde a$,  the fermionic partner of the axion,
can change dramatically the property of dark matter.

First of all, the axino can be the lightsest supersymmetric particle (LSP) in $R$-parity conserving models. The axino in the KSVZ model decouples at a very high temperature and thus it must be very light ($m_{\tilde a} \sim 0.1$ keV) to become a viable dark matter candidate \cite{Raja91}.
In the gravity mediated supersymmetry breaking scheme, such a light axino mass could be arranged although it is generic to have the axino mass of order of the gravitino mass, $10^2-10^3$ GeV \cite{Goto92,Chun92,Chun95}.
In the DFSZ model or in the KSVZ model with lower reheat temperature, the axino may never be in thermal equilibrium, but thermal regeneration (the axino production from the scattering, decay and inverse decay of thermal particles) becomes an important source of the cosmic axino abundance
\cite{Chun00,Covi01,Brand04,Strumia10,Chun11,Bae11,KYChoi11}.
Although suppressed by the intermediate PQ symmetry breaking scale,
$v_{PQ}$, the axino interactions turn out to be efficient enough to generate a huge
number of axinos unless the reheat temperature is much lower than the weak scale.

If the axino has a typical mass in the range of $10^2-10^3$ GeV and the reheat temperature is above the electroweak scale,
the axino cannot be dark matter
as its thermal population easily overcloses the Universe. Then, it has to decay into the LSP dark matter such as the usual neutralino LSP \cite{KYChoi08,Baer11,Chun11} or gravitino \cite{Cheung11}.  Since the axino lifetime is suppressed by $v_{PQ}^2$, the axino decay may occur very late to produce the neutralino dark matter after its freeze-out causing overabundant neutralino density.  This problem can be evaded if $v_{PQ}$ takes its lower value to render the axino decay before the neutralino freeze-out, otherwise the neutralino LSP must have a large annihilation
cross-section to allow a sufficient reannihilation \cite{KYChoi08}.

\medskip

The thermal axino production in the KSVZ model has been studied extensively
taking a simple assumption of the effective QCD interaction below the PQ scale \cite{Covi01,Brand04,Strumia10}
and more carefully considering a general situation \cite{Bae11}.
Axino  couplings in the DFSZ model are different from those in the KSVZ model and
their impact on the thermal axino production has been considered recently
in Refs.~\cite{Chun11,Bae11,KYChoi11}.
In this paper, we provide more general analysis of the DFSZ axino abundance including the scattering, decay and inverse decay processes.  Then, we explore the Bino-Higgsino dark matter parameter region depending on the PQ symmetry breaking scale.

The paper is organized as follows.
In Section 2, we describe the supersymmetric DFSZ model to derive the dominant axino couplings.
In Section 3, the DFSZ axino abundances coming from various channels are calculated
and compared with each other.  Since  the DFSZ axinos are overproduced as far as the reheat temperature is above the weak scale, we investigate its impact on the neutralino dark matter property by taking some benchmark parameter points in Section 4.  We conclude in Section 5.

\section{Axino couplings in the supersymmtric DFSZ model}

One of the hierarchy problems in the minimal supersymmetric standard model is the $\mu$ problem:
``why the Higgs bilinear parameter, $\mu$, sits at the TeV scale?". This problem can be nicely resolved in connection with the axion solution of the strong CP problem
\`a la Kim-Nilles \cite{Kim84}.  Extending the Higgs sector with
\begin{equation}
 W_{DFSZ} = \lambda_H {P^2 \over M_P} H_u H_d,
\end{equation}
where $M_P$ is the reduced Planck mass,
one can introduce the PQ symmetry under which the Higgs doublet superfields $H_{u,d}$ and the gauge
singlet field $P$ are charged. In order to break the PQ symmetry, one can introduce a PQ sector like
\begin{equation} \label{PQ-potential}
 W_{PQ} = \lambda_S S\biggl(PQ-\frac{v_{PQ}^2}{2}\biggr)
\end{equation}
which will lead to $\langle P \rangle \sim \langle Q \rangle \sim v_{PQ}/\sqrt{2}$.
One can then get an appropriate $\mu$ term: $\mu = \lambda_H \langle P \rangle^2/M_P$.
This is nothing but the supersymmetric realization of the DFSZ axion model \cite{dfsz}.
In this scheme, the axion supermultiplet is a combination of mostly $P$ and $Q$, and
a small mixture of $H_u$ and $H_d$ \cite{Chun95}. Thus, the axino interaction is given by
the axino-Higgs-Higgsino:
\begin{equation} \label{dfsz-coupling}
 {\cal L}_{DFSZ} = c_H {\mu \over v_{PQ}}  \tilde a\, [H_u \tilde H_d + \tilde H_u H_d] + h.c.
\end{equation}
And after the electroweak symmetry breaking, the axino-top-stop coupling induced by the axino-Higgsino mixing also becomes important:
\begin{equation} \label{dfsz_top-coupling}
 {\cal L}_{DFSZ, top}=c_t {m_t\over v_{PQ}} \tilde a\, [ t \tilde t^c + \tilde t t^c] + h.c.
 \end{equation}
where $c_H$ and $c_t$ are order-one parameters depending on the PQ symmetry breaking sector and other couplings proportional to smaller quark or lepton masses are neglected. As the axino couplings are of order $\mu/v_{PQ} \lesssim 10^{-8}$, the axinos cannot be in thermal equilibrium but can be produced thermally by the scattering, decay and inverse decay coming from the couplings (\ref{dfsz-coupling}) and (\ref{dfsz_top-coupling}). For our numerical calculation of the thermal axino production in the next section, we will put
$c_H=2$ and $c_t=2$, which corresponds to assigning the PQ charge 1 to each Higgs supermultiplet, respectively, with the PQ sector Eq. (\ref{PQ-potential}).\footnote{Note that the physically relevant parameter is $c_{H,t}/v_{PQ}$ so that $c_{H,t}$ also depends on the normalization convention on $v_{PQ}$. $c_t$ can be obtained from the axino-Higgsino mixing Eq. (\ref{axino-higgsino}) with $\mu \gg m_{\tilde a}$.}
Note that the axino-top-stop coupling arises only after the electroweak
symmetry breaking and thus does not contribute to the thermal axino production if the temperature of the Universe is larger than the weak scale.

\medskip

Let us now compare the DFSZ axino couplings to the KSVZ axino couplings.
In the supersymmetic KSVZ model, one introduces a heavy quark pair $(X,X^c)$ carrying the PQ charge:
\begin{equation}
 W_{KSVZ} = \lambda_X P X X^c
\end{equation}
from which the heavy quark gets an intermediate mass, $M_X \sim \lambda_X  v_{PQ}/\sqrt{2}$. 
Below the scale $M_X$, the axino gets an effective QCD interaction:
\begin{equation} \label{taGG}
 {\cal L}_{QCD} = {g_s^2\over 32\pi^2}
 {1\over \sqrt{2}v_{PQ} }\, \tilde{a}\, \sigma^{\mu\nu} \tilde g^a
 G^a_{\mu\nu} + h.c. \,,
\end{equation}
from which the previous calculation of the KSVZ axino abundance was preformed \cite{Covi01,Brand04,Strumia10} to show that the axino relic density is proportional to the reheat temperature $T_R$ of the Universe. However, a recent study showed that such a $T_R$--dependence does
not occur if $T_R$ is between $v_{PQ}$ and $M_X$ as the axino abundance is determined
by the Yukawa coupling $\lambda_X \sim \sqrt{2}M_X/v_{PQ}$ and this property is applied to 
the DFSZ axino \cite{Bae11}.\footnote{Note that a different claim was made in Ref.~\cite{KYChoi11}.}

\begin{figure}
\subfigure[]{\includegraphics[width=7cm]{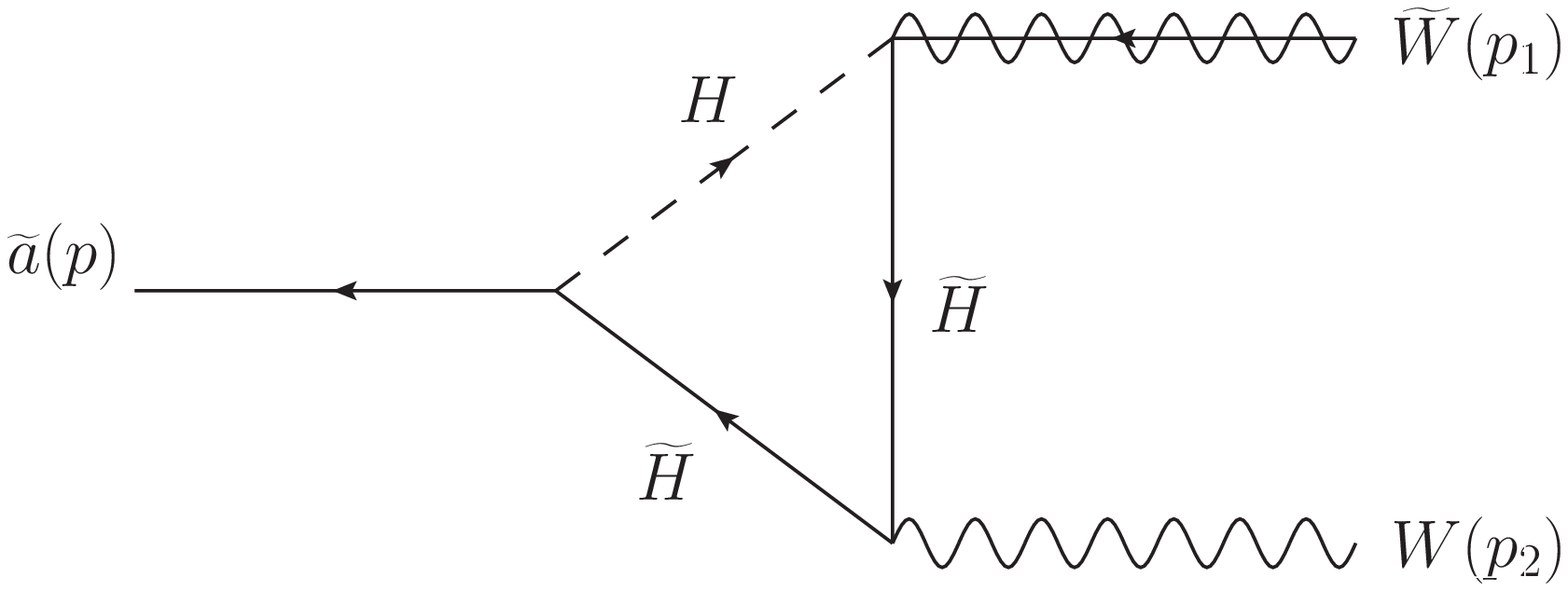}}~
\subfigure[]{\includegraphics[width=7cm]{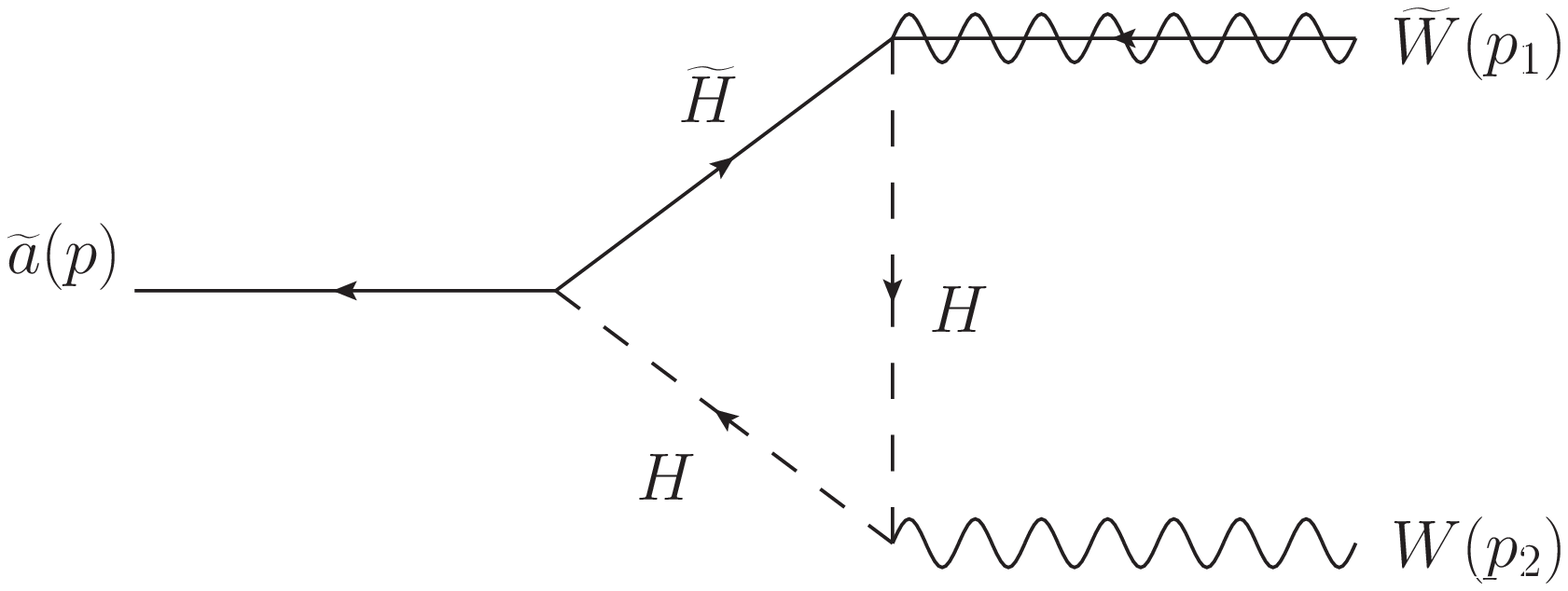}}
\caption{\label{feyn:loop}
1PI axino-gaugino-gauge boson interaction at one-loop level for the DFSZ axino}
\end{figure}

Let us recapitulate the result of Ref.~\cite{Bae11} to show that the DFSZ axino abundance is 
independent of the reheat temperature.  After the PQ symmetry breaking, we get the effective 
axino coupling with the Higgs and Higgsino fields (\ref{dfsz-coupling}) (which corresponds to the 
non-linear realization of the PQ symmetry with the parameter $c=0$ in Ref.~\cite{Bae11}).
This leads to the 1PI axino-gaugino-gauge boson interaction of dimension 5 from the diagram in Fig.~\ref{feyn:loop}. The corresponding amplitude is given by
\begin{equation} \label{axinoWW}
{\cal A} = {g_2^2\over 16\pi^2\sqrt{2} v_{PQ}} c_H 
\int^1_0 dx \int^{1-x}_0 dy 
\frac{2\mu^2 \bar{u}(p_1+p_2)\sigma_{\mu\nu} \gamma_5 u(p_1)\epsilon^\mu p_2^\nu }{\mu^2-[p_1^2x(1-x)+p_2^2y(1-y) +2(p_1 \cdot p_2)xy]}
\end{equation}
for the axino and the $SU(2)$ gauge supermulitiplet interaction.  It is then easy to see the limiting behavior of the loop integral: $1/p^2$ if $|p| \gg \mu$, and $1/\mu^2$ if $\mu \gg |p|$ assuming $\mu \sim m_H$, which tells us that the axino production through 
the effective axion-gaugino-gauge boson interaction is suppressed  by 
$\mu^2/T^2$ for $T > \mu$
if there is no other heavier
gauge charged and PQ charged supermultiplet than the scale $\mu$, which is the case for the ``pure" DFSZ models.\footnote{We refer the ``pure"
DFSZ to the models where gauge charged and PQ charged particles are only among the MSSM particles, and the ``mixed" DFSZ to the models containing exotic gauge charged and PQ charged supermultiplets in addition to the MSSM particles.} In this case, the axino production
will be determined by the tree level interaction Eq.~(\ref{dfsz-coupling}) so that the DFSZ axino abundance becomes independent of $T_R$. Still if we assume some additional heavier gauge charged and PQ charged supermultiplet of some mass $M$ larger than the scale $\mu$,
the axino production will be dominated by the effective interaction Eq.~(\ref{taGG}) between the energy scales $8\pi^2\mu$ and $M$ as can be seen in the table 1 of \cite{Bae11},
and the axino abundance becomes proportional to $T_R$ in this regime as in the KSVZ case.
Previous studies on the DFSZ axino production with
the Eq.~(\ref{taGG}) as the dominant interaction at high reheat temperature effectively correspond to such ``mixed" DFSZ models.
In this paper, we  consider the ``pure" DFSZ case with the Higgs supermultiplet
as the heaviest gauge charged and PQ charged supermultiplet.
So while the DFSZ axino abundance is independent of $T_R$,
the KSVZ axino abundance is proportional to $T_R$ and can be larger than the DFSZ axino abundance
for larger $T_R$ \cite{Chun11}.
 
Note that the axino effective interactions with $SU(2)\times U(1)$ gauge supermultiplets are  obtained below the scale $\mu$ by integrating out the Higgs supermultiplets.
In a similar way, the effective axino-gluino-gluon interaction (\ref{taGG}) in the ``pure'' DFSZ model arises after integrating out the quark supermultiplet having the Yukawa coupling as in Eq.~(\ref{dfsz_top-coupling}), along with its supersymmetric counterpart: the usual effective axion-gluon-gluon interaction solving the strong CP problem.

\medskip

Another distinction of the DFSZ axino from the KSVZ axino comes from the lifetime calculated from
the couplings (\ref{dfsz-coupling}) and  (\ref{taGG}), respectively. As was shown in Ref.~\cite{KYChoi08}, the thermal abundance of the KSVZ axino can dominate the energy density of the Universe for larger $v_{PQ}$.  But the DFSZ axino lives shorter than the KSVZ axino and thus it always decays before it overdominates the Universe as will be discussed in Section \ref{axino_decay}.

\section{Thermal production of the DFSZ axino}

\begin{figure}
\subfigure[]{\includegraphics[width=5cm]{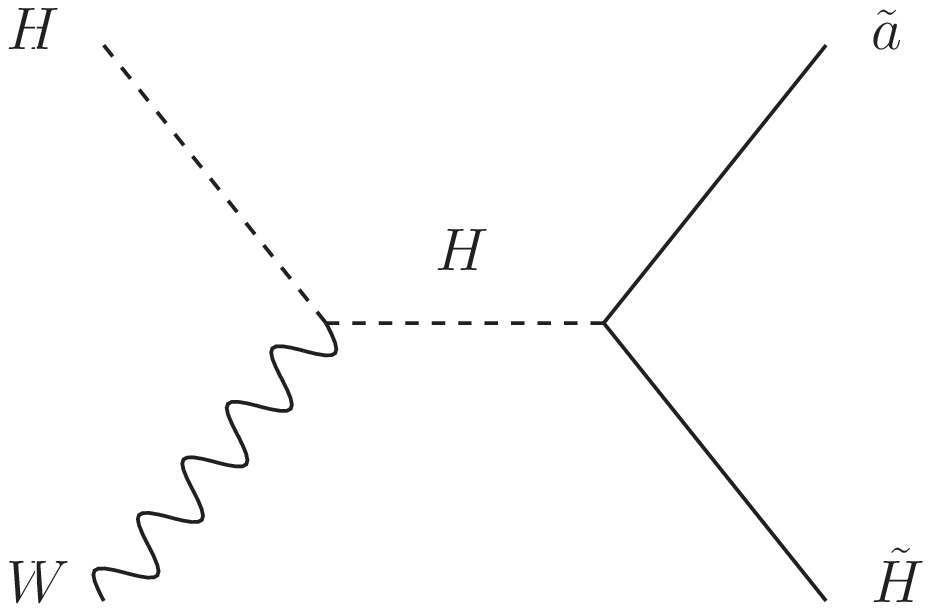}}~
\subfigure[]{\includegraphics[width=5cm]{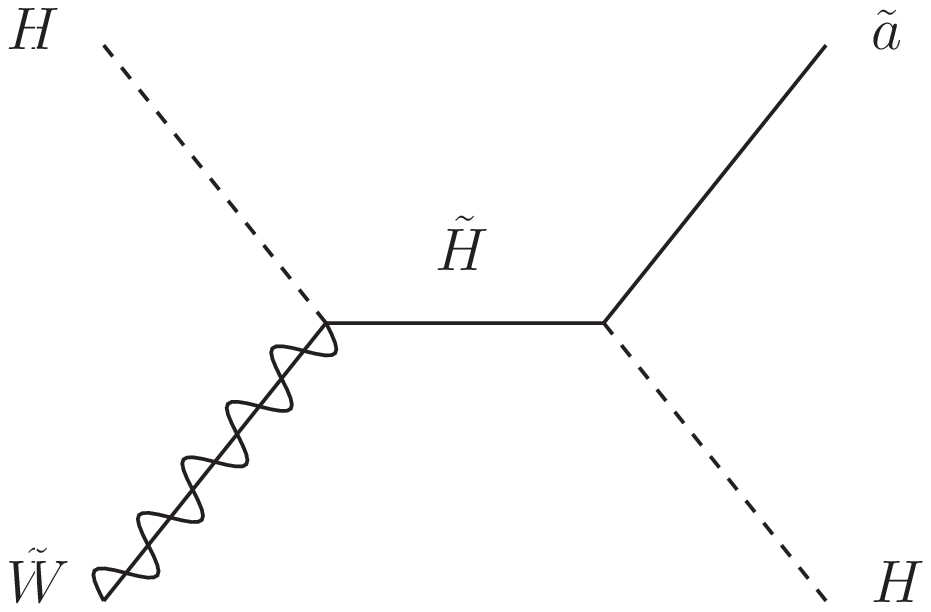}}
\caption{\label{feyn:gauge}
Examples of scattering diagrams (considered in Fig.~3-7 of Ref.~\cite{Bae11}) involving the axino-Higgs-Higgsino coupling and gauge couplings of the Higgs multiplet. There are also two other diagrams with supersymmetric gauge vertices and t, u-channel diagrams. All
processes are included in the calculation.}
\end{figure}

\begin{figure}
\subfigure[]{\includegraphics[width=5cm]{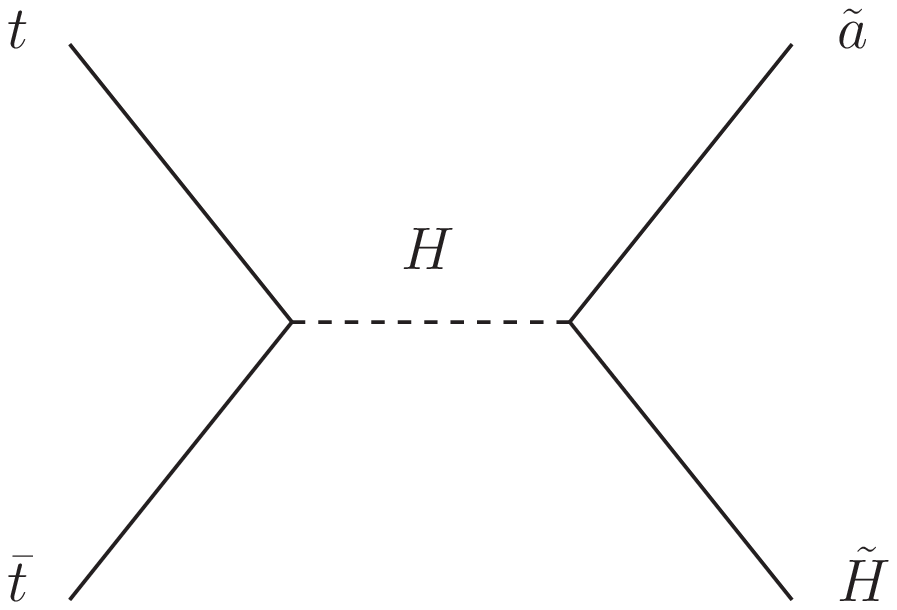}}~
\subfigure[]{\includegraphics[width=5cm]{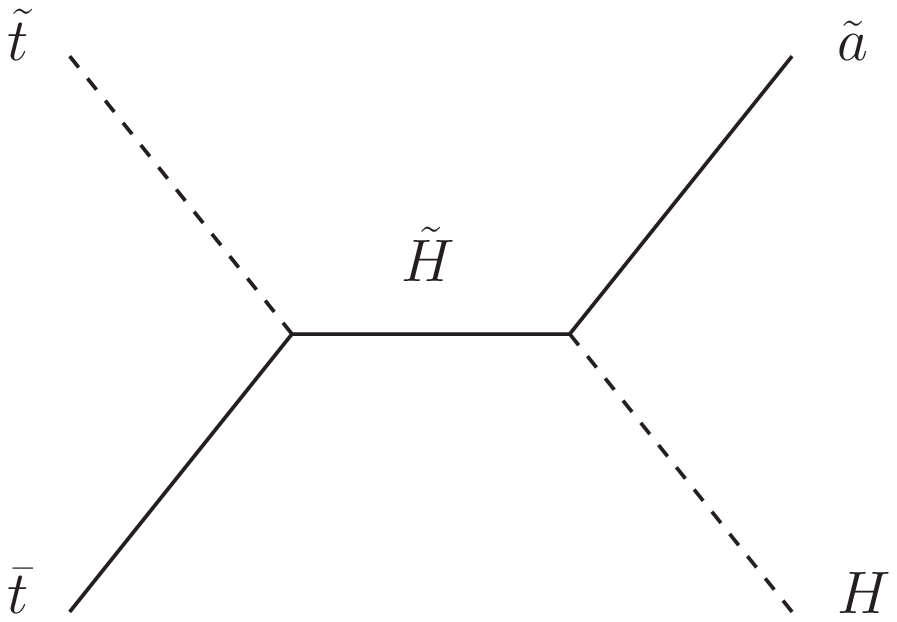}}~
\subfigure[]{\includegraphics[width=5cm]{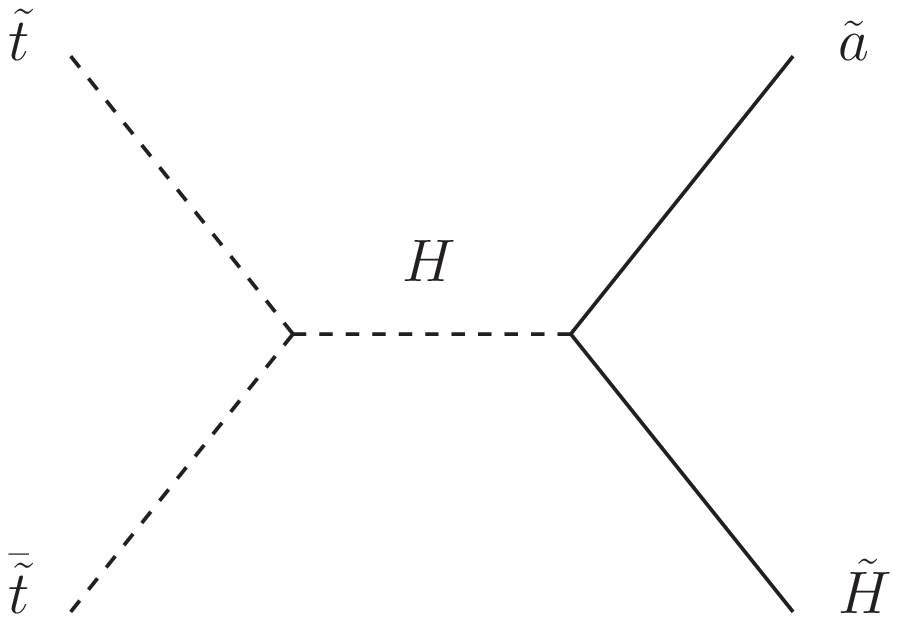}}
\caption{\label{feyn:top} 
Examples of scattering diagrams involving the axino-Higgs-Higgsino coupling and top Yukawa coupling. There are also t, u-channel
diagrams. All processes are included in the calculation.}
\end{figure}

Thermal production of the DFSZ axino has been computed previously
considering the decay and inverse decay in Ref.~\cite{Chun11} and
the scattering involving the axino-Higgsino-Higgs coupling and gauge coupling of Higgs multiplet in Ref.~\cite{Bae11} as depicted in Fig.~\ref{feyn:gauge}.
There are additional processes involving the top Yukawa coupling as depicted in Fig.~\ref{feyn:top} which can be potentially more important since the top Yukawa coupling is of order of unity.
In this section, we combine all the possible processes to evaluate the axino yield for each channel.
For the calculation, we take some benchmark points for the relevant particle masses
including supersymmetry breaking effect  both in the unbroken phase and in the broken phase of the electroweak symmetry.

\begin{figure}
\includegraphics[width=10cm]{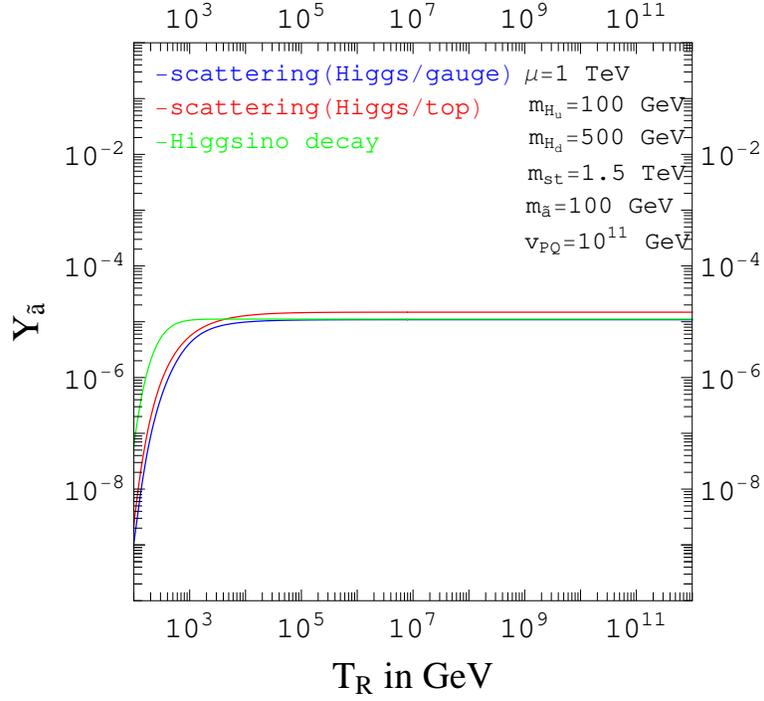}
\caption{\label{fig:DFSZ1}
The axino yield from the scattering diagrams in Fig.~\ref{feyn:gauge}  (blue line), the scattering  diagrams in Fig.~\ref{feyn:top}  (red line), and the decay $\tilde H \to H \tilde a$ (green line). }
\end{figure}

\begin{figure}
\includegraphics[width=10cm]{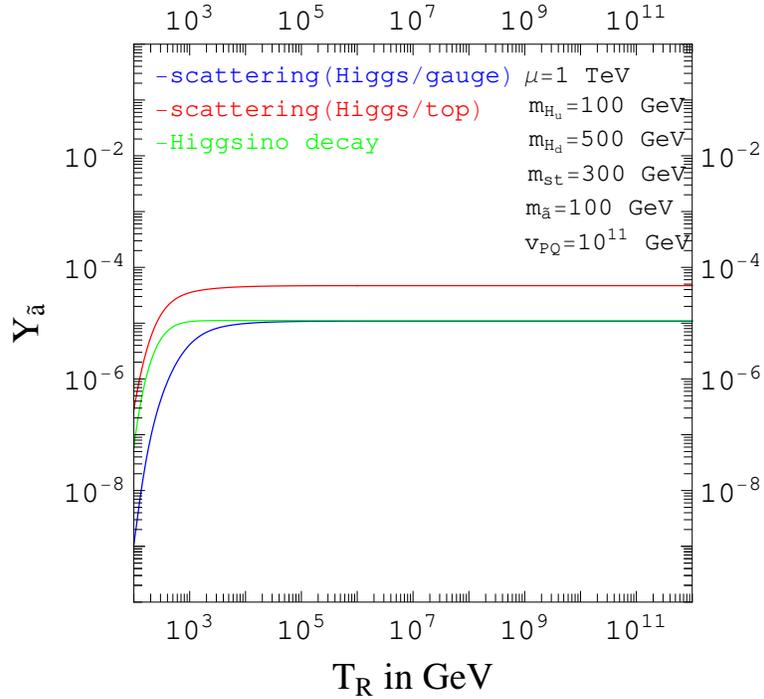}
\caption{\label{fig:DFSZ_resonance}
Same as in Fig.~\ref{fig:DFSZ1} but with the parameter choice allowing a resonant axino production
in the channel of Fig.~\ref{feyn:top}(b) }
\end{figure}

\begin{figure}
\includegraphics[width=10cm]{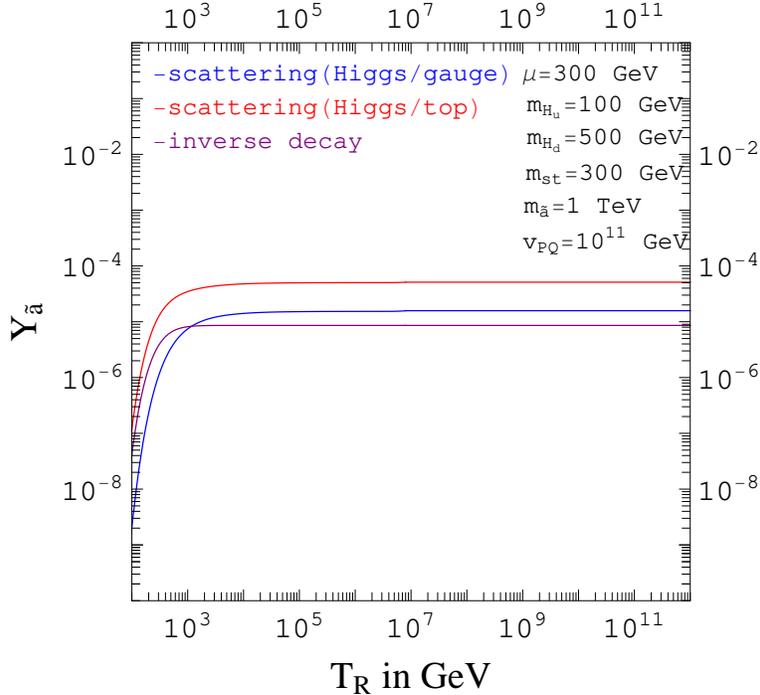}
\caption{\label{fig:DFSZ_TeV}
Same as in Fig.~\ref{fig:DFSZ_resonance} but with parameter choice allowing the inverse decay  $\tilde{H}H\to\tilde{a}$ (purple line)}
\end{figure}

\begin{figure}
\includegraphics[width=13cm]{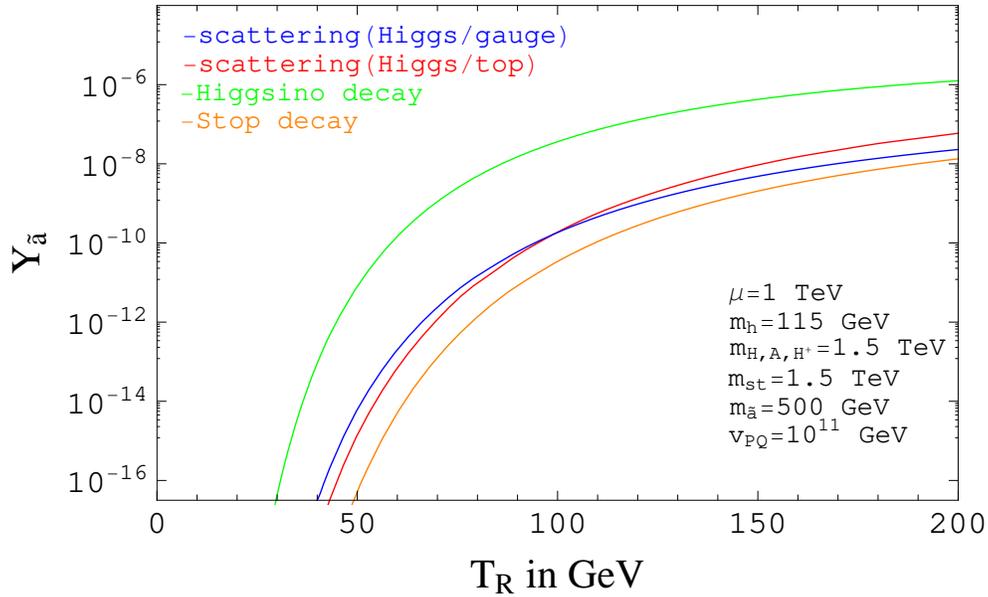}
\caption{\label{fig:DFSZ_ew}
The axino yield at low reheat temperature below the electroweak scale with a decoupled benchmark spectrum}
\end{figure}

In Fig.~\ref{fig:DFSZ1}, the thermal axino number density $Y_{\tilde a} \equiv n_{\tilde a}/s$
per the entropy density is analyzed for the benchmark points:
$\mu=1$ TeV, $m_{H_u}=100$ GeV, $m_{H_d}=500$ GeV,\footnote{$m_{H_u}$ and $m_{H_d}$ denote $(m_2^2+|\mu|^2)^{1/2}$ and $(m_1^2+|\mu|^2)^{1/2}$, where $m_2$ and $m_1$ are soft mass terms of $H_u$ and $H_d$, respectively} $m_{\tilde{t}}=1.5$ TeV, and
$m_{\tilde{a}}=100$ GeV with $v_{PQ}=10^{11}$ GeV. In this parameter space, the Higgsino decay
$\tilde H \to H \tilde a$ is allowed to contribute to the thermal axino production in addition to
the scattering processes from Fig.~\ref{feyn:gauge} and \ref{feyn:top} (red line).
As can be seen in the plot, three contributions are almost the same
in the region of $T_R\gtrsim\mu$.
In the slightly lower temperature, i.e. $T_R\lesssim\mu$, the decay contribution is
about an order of magnitude larger than the others because scattering processes are doubly suppressed by the Boltzmann factor while decay process has only one Boltzmann suppression.

The second example is depicted in Fig.~\ref{fig:DFSZ_resonance} for which we take
the benchmark points: $\mu=1$ TeV, $m_{H_u}=100$ GeV, $m_{H_d}=500$ GeV, $m_{\tilde{t}}=300$ GeV, $m_{\tilde{a}}=100$ GeV with $v_{PQ}=10^{11}$ GeV.
The only difference from the previous example is the stop mass. In this example,
a small stop mass is considered to allow a resonant axino production in the channel $\tilde{t}+\bar{t}\to\widetilde{H}\to\tilde{a}+H$ (Fig.~\ref{feyn:top}(b)).
The resonant process can enhance the production cross section by factor of $1/\Gamma^2$, where $\Gamma$ is the decay rate of the intermediate particle, the Higgsino in this case.
As the axino thermal production is obtained by integrating the cross section over $T=[0,T_R]$,
such a resonance effect is averaged out by the width of the resonance, $\Gamma$, resulting in the
enhancement of the axino yield by factor of $1/\Gamma$.
In this example, we have  a quite large decay rate of Higgsino, $\Gamma\sim100$ GeV, leading to $\mu/\Gamma\sim{\cal O}(10)$ enhacement for the resonant production channel.
Thus, the corresponding scattering process dominates the others by about factor 4.

The third case, shown in Fig.~\ref{fig:DFSZ_TeV},\footnote{Our results in Figs.~\ref{fig:DFSZ1}--\ref{fig:DFSZ_TeV} are different from Fig.~3 of Ref.~\cite{KYChoi11} showing $T_R$ dependence as remarked in Section II.}
 is for a heavy axino which can decay to a Higgs boson and a Higgsino:  $m_{\tilde{a}}=1$ TeV and $\mu=300$ GeV with the other parameters kept same as in the previous case. As one can see, the axino yield from the inverse decay is more or less the same as the decay contribution in the previous examples.

From the above calculation, one finds the axino yield $Y_{\tilde a}\sim 10^{-5} (10^{11} \mbox{ GeV}/v_{PQ})^2$ for each allowed channel of scattering, decay or inverse decay
as far as the reheat temperature above the electorweak scale ($T_R \gtrsim \mu$).
In some cases, a resonant production can occur to enhance the axino yield in a certain
scattering process. As a consequence, thermally produced axinos would severely
overclose the Universe if they take a weak scale mass and are stable.

\medskip

The thermal axino production can be effectively suppressed if the reheat temperature
of the Universe is below the weak scale. In such a case,
the relevant processes for axino production strongly depend on
a specific mass spectrum with broken electroweak symmetry.\footnote{
 In this calculation, we use the zero temperature value for the vacuum expectation value of the Higgs field $v=174$ GeV.}
For the purpose of our presentation, we will discuss only a
representative behavior taking a simplified benchmark point.
To this end, we assume that the stop, Wino and CP-odd Higgs are heavy enough ($\sim$ 1.5 TeV) to be decoupled at low reheat temperature. Also we set $m_{\tilde{a}}=500$ GeV
which is taken to be the reference value in the next section discussing heavy axino decay.
For the Higgsino mass, we choose $\mu=1$ TeV allowing the  decay $\tilde{H} \to h \tilde{a}$.
Note also that the $\tilde{t} \to t \tilde{a}$ decay processes are possible.
Fig.~\ref{fig:DFSZ_ew} shows the result for the axino yield in this benchmark
point at low reheat temperature below 200 GeV.
As we have observed in Fig.~\ref{fig:DFSZ1},
the decay contribution from the Higgsino is dominant over the scattering contributions
because of less Boltzmann suppression.
On the other hand, the contribution from the stop decay is smaller because the process becomes available from the Higgsino-axino mixing by the electroweak symmetry breaking and thus involves
the top mass $m_t$ which is much smaller than $\mu$ appearing in the Higgsino decay process. We can observe that all contributions become rapidly decreasing after the electroweak scale and drop down drastically below 50 GeV being heavily suppressed by the Boltzmann factors.
Due to such a high Boltzmann suppression, the heavy axino can have a right dark matter density
for $T_R \sim 50$ GeV.

\section{Heavy Axino Decay and Neutralino Dark Matter \label{axino_decay}}

In this section, we explore the cosmological impact of a heavy axino which cannot
be the LSP and thus has to decay to a dark matter candidate which we assume to be a neutralino
LSP.  Such a scenario has been considered first in Ref.~\cite{KYChoi08} for the case of
the KSVZ axino and later in Ref.~\cite{Chun11} for the case of the DFSZ axino.
Generalizing the latter analysis, we examine the dark matter property depending on the neutralino parameter space and the PQ scale.

Depending on the mass spectrum, the axino can decay to a Higgsino and a Higgs, a Neutralino (Bino-like) and a photon, etc. We assume here, for simplicity and consistency with current experiment \cite{Aad:2011ib,Chatrchyan:2011zy}, that all the squarks and sleptons are heavier
than the Higgsino. One can consider two regimes of the axino mass as follows.
\begin{itemize}
\item For $m_{\tilde{a}}>\mu$, the dominant decay process of the axino is
\begin{equation}
\tilde{a}\to \widetilde{H}^0+H^0/Z\qquad \text{or}\qquad \tilde{H}^{\pm}+H^{\mp}/W^{\mp}
\end{equation}
where $H^0$ stands for neutral Higgs states, i.e. $h$, $H$, or $A$.
This decay mode is due to the tree-level axino-Higgs-Higgsino coupling.
If allowed, it will provide the most efficient axino decay.
\item For $m_{\tilde{a}}<\mu$, possible decay processes are
\begin{eqnarray}
&&\tilde{a}\to \tilde{\chi}+\gamma\label{axino_decay:chi_gamma}\\
&&\tilde{a}\to H^0/H^{\mp}+\widetilde{H}^{0*}/\widetilde{H}^{\pm*}\to H^0/H^{\mp}+H^0/Z/H^{\mp}/W^{\mp}+\widetilde{\chi}\label{axino_decay:3body}\\
&&\tilde{a}\to H^0+\tilde{\chi}\label{axino_decay:H_chi}\\
&&\tilde{a}\to Z+\tilde{\chi}\label{axino_decay:Z_chi}
\end{eqnarray}
The first mode is due to the 1PI axino-photon-neutralino interaction. The second mode is a 3-body decay due to the tree-level axino-Higgs-Higgsino interaction.  The third mode is a tree-level process due to the axino-Higgs-Higgsino coupling with the Higgsino-Bino and axino-Higgsino mixing. Finally the last one is a tree-level process due to the axino-Higgsino mixing.
\end{itemize}

\begin{figure}
\includegraphics[width=14cm]{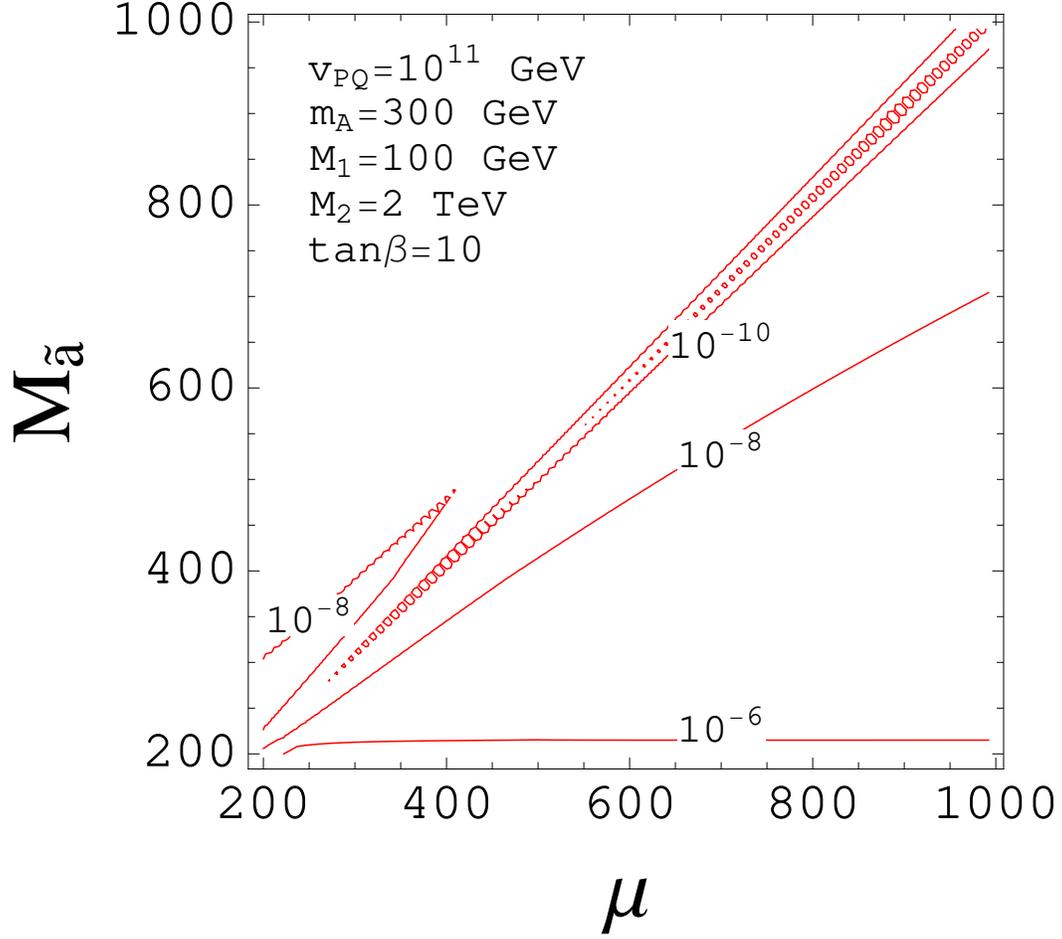}
\caption{Contour plot of the axino lifetime (sec). \label{fig:axino_decay_1e11}}
\end{figure}

Following the calculation described in Appendix,
we compute the total decay rate in Fig.~\ref{fig:axino_decay_1e11}.
Unlike the KSVZ case where the axino yield depends on the reheat termperature, the DFSZ axino yield is determined by the mass spectrum and can be parameterized by
\begin{equation}
Y_{\tilde{a}}=10^{-5}\xi\biggl(\frac{\mu}{\text{TeV}}\biggr)^2\biggl(\frac{10^{11}\text{ GeV}}{v_{PQ}}\biggr)^2
\end{equation}
if the reheat temperature is larger than the electroweak scale.
Here $\xi$ is an order-one quantity depending on the particle mass spectrum.
Such abundant axinos will decay to yield overproduction of neutralinos which may thermalize or
reannihilate depending on the axino lifetime.
The axino decay temperature $T_D$ is given by
\begin{equation}
T_D=1.4 \text{ GeV}\biggl(\frac{70}{g_*}\biggr)^{1/4}\biggl(\frac{3\times10^{-7}\text{ s}}{\tau_{\tilde{a}}}\biggr)^{1/2}.\label{eq:T_D}
\end{equation}
If $T_D$ is larger than the freeze-out temperature  $T_{f}$ of the neutralino LSP,
the overproduced LSPs
will thermalize to be settled down to the usual freeze-out relic density.
For $T_D<T_{f}$, the overabundant neutralino density can be depleted by reannhilation and
the final relic density is determined by \cite{KYChoi08}
\begin{equation} \label{re-ann}
Y_{\tilde{\chi}}^{-1}(T)\approx Y_{\tilde{\chi}}^{-1}(T_D)+\frac{\langle \sigma_Av\rangle s(T_D)}{H(T_D)},
\end{equation}
where $Y_{\tilde{\chi}}(T_D)=Y_{\tilde{\chi}}(T_{\text{fr}})+Y_{\tilde{a}}\approx Y_{\tilde{a}}$
for the reheat temperature above the electroweak scale. Since $T_D<T_f$, we need the annihilation rate $\langle \sigma_Av\rangle$ of the neutralino LSP
larger than the canonical value in order to get a right relic density for dark matter.

For low reheat temperature, $Y_{\tilde{a}}$ can become comparable to or smaller than $Y_{\tilde{\chi}}(T_{\text{fr}})$.
For example, we can find this occurs around the temperature $T \sim$ 40 GeV, 50 GeV, and 60 GeV for $v_{PQ} = 10^{10}$ GeV, $10^{11}$ GeV, and $10^{12}$ GeV, respectively, from the result of Fig.~\ref{fig:DFSZ_ew} with the assumed mass spectrum.
Below this temperature, the neutralino relic abudance is given by the standard results
with a negligible axino contribution.

In the case of the KSVZ axino, the axino lifetime determined by the coupling (\ref{taGG}) can be
long enough so that there could occur a period of axino matter domination \cite{KYChoi08}.
On the other hand, the DFSZ axino lifetime determined by the coupling (\ref{dfsz-coupling}) is typically shorter
than that of the KSVZ axino, and thus there occurs no DFSZ axino domination as
shown explicitly in Figs.~\ref{fig:1e10_M2_2000}-\ref{fig:1e12_M2_2000}.
To get an rough estimate, let us consider the largest axino lifetime $\sim 10^{-5}$ sec found
in Fig.~\ref{fig:1e12_M2_2000} for $v_{PQ}=10^{12}$ GeV.
This give $T_D \sim 0.1$ GeV from Eq.~(\ref{eq:T_D}). This can be compared with the axino-radiation equality temperature $T_{eq}$
determined by $T_{eq}=4m_{\tilde{a}}Y_{\tilde{a}}/3$ \cite{KYChoi08} which give
$T_{eq} \sim 10^{-4}$ GeV for $m_{\tilde a} \sim 1$ TeV and $Y_{\tilde a} \sim 10^{-7}$. Thus, it becomes clear to have $T_{eq} < T_D$,
that is, the axino decays before it overdominates the Universe, for almost all the
parameter space.
Note that  the axino lifetime can become much larger near the boundary line between the neutralino LSP and axino LSP, so the axino dominated universe can be realized. In this case, however, the neutralino abundance becomes too large to get right dark matter density.

\begin{figure}
\includegraphics[width=8cm]{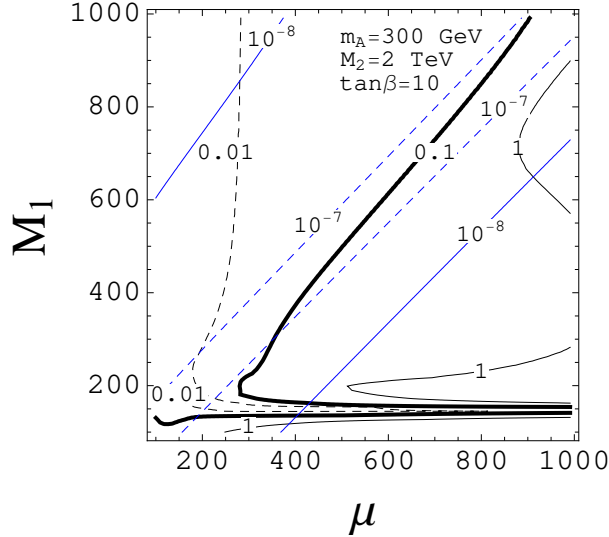}
\caption{Black lines stand for the standard freeze-out neutralino abundance. Blue lines stand for spin-independent direct detection cross section with proton in unit of pb. The
WMAP dark matter abundance is depicted by thick line.
\label{fig:Oh2_MSSM_M2_2000} }
\end{figure}
\begin{figure}
\includegraphics[width=8cm]{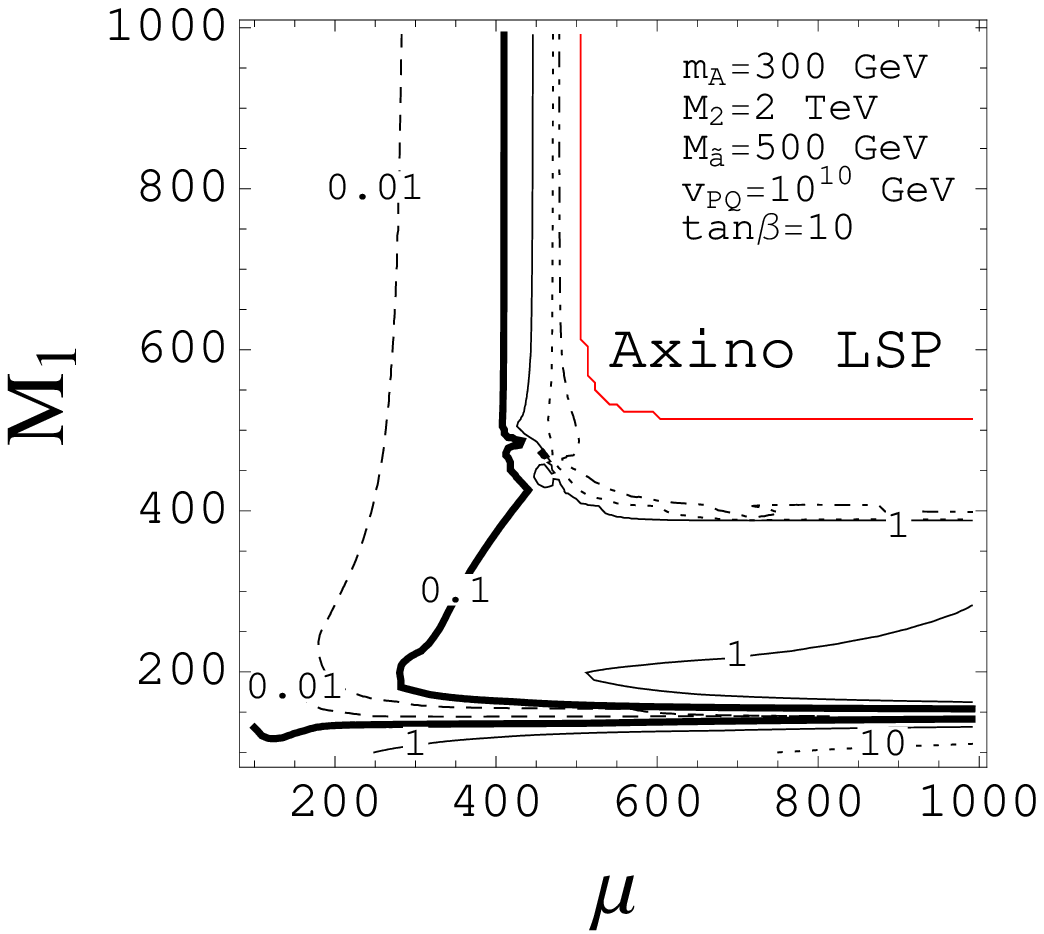}\quad
\includegraphics[width=8cm]{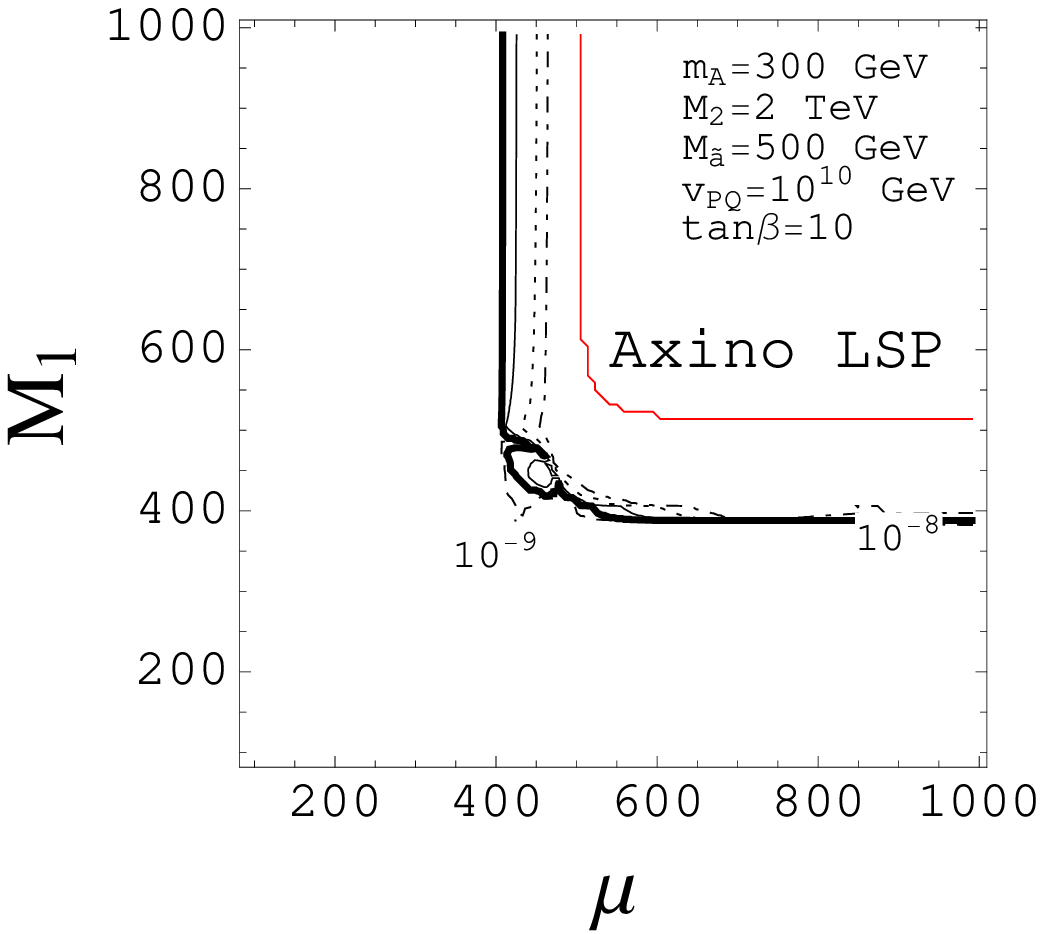}
\caption{The left panel shows the neutralino relic abundance after the axino decay, and the right
panel shows the axino lifetime for $v_{PQ}=10^{10}$ GeV.  Red line is the border line between the axino LSP and the neutralino LSP.
\label{fig:1e10_M2_2000}}
\end{figure}
\begin{figure}
\includegraphics[width=8cm]{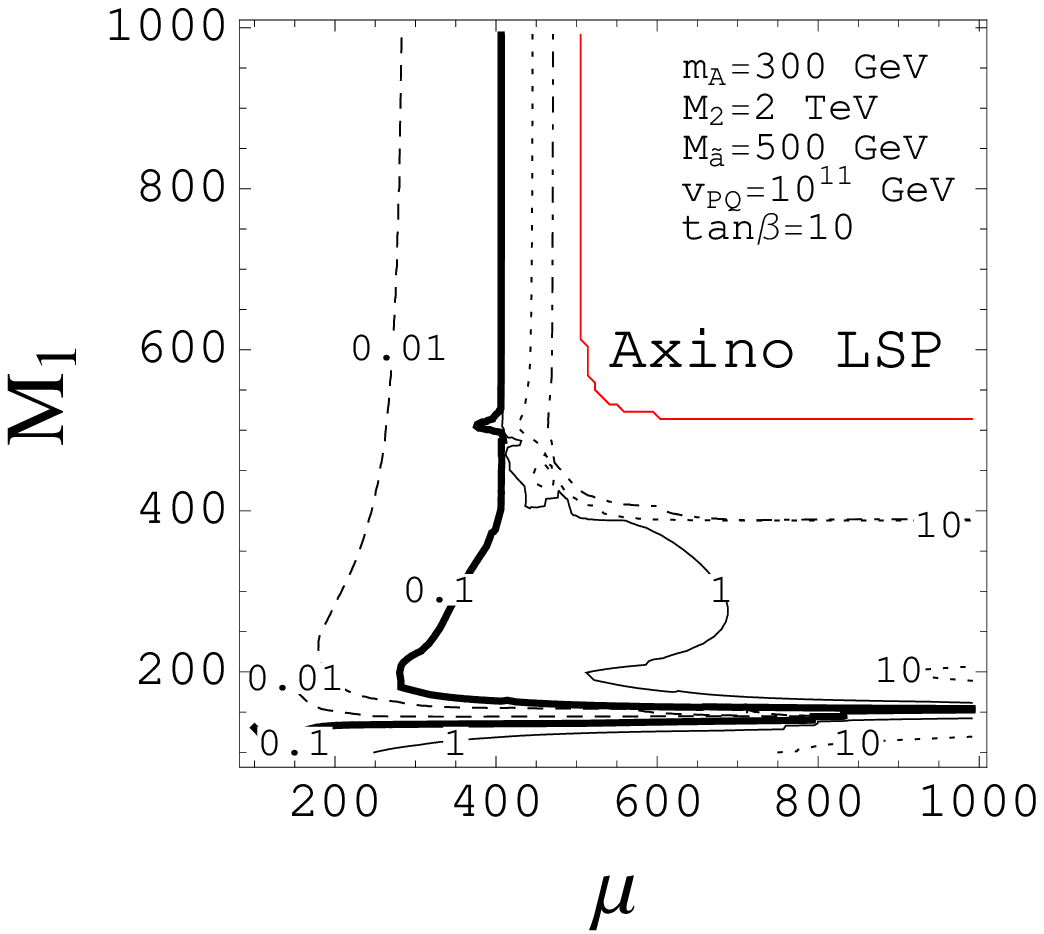}\quad
\includegraphics[width=8cm]{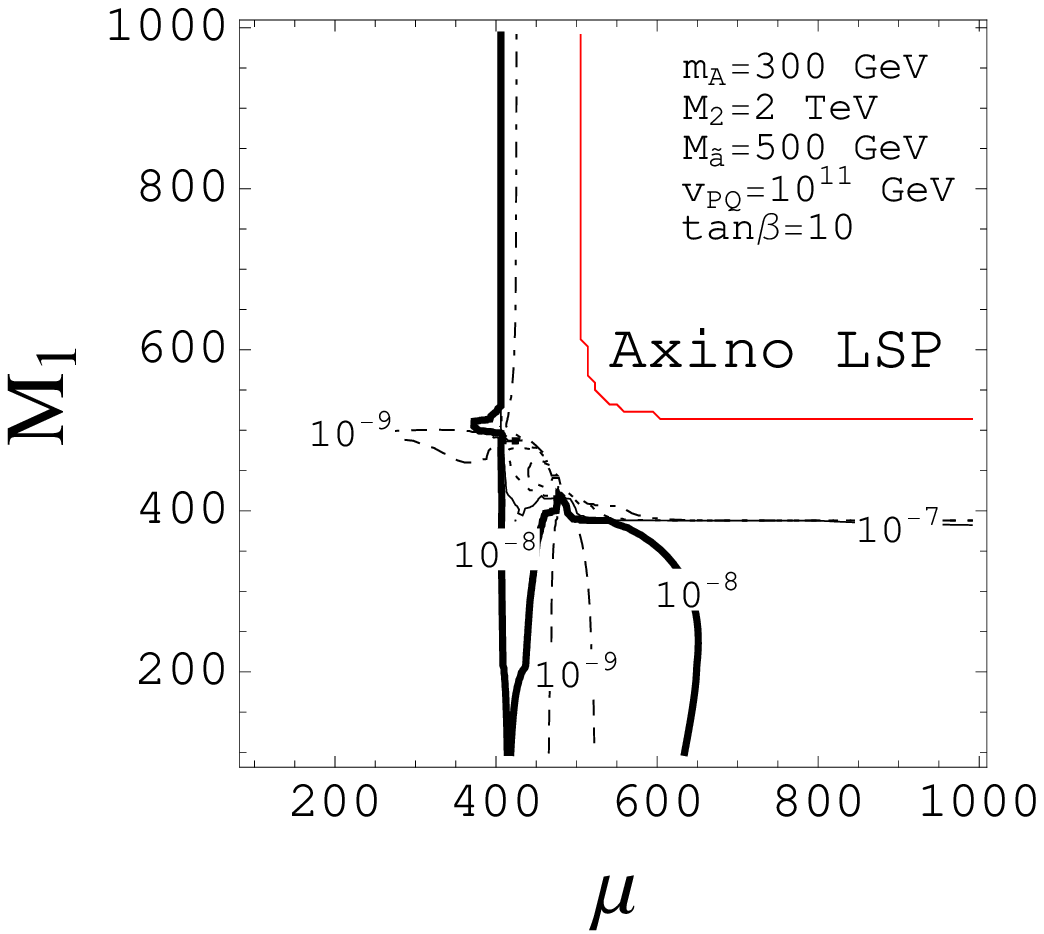}
\caption{Same as the previous figure with $v_{PQ}=10^{11}$ GeV.
\label{fig:1e11_M2_2000}}
\end{figure}
\begin{figure}
\includegraphics[width=8cm]{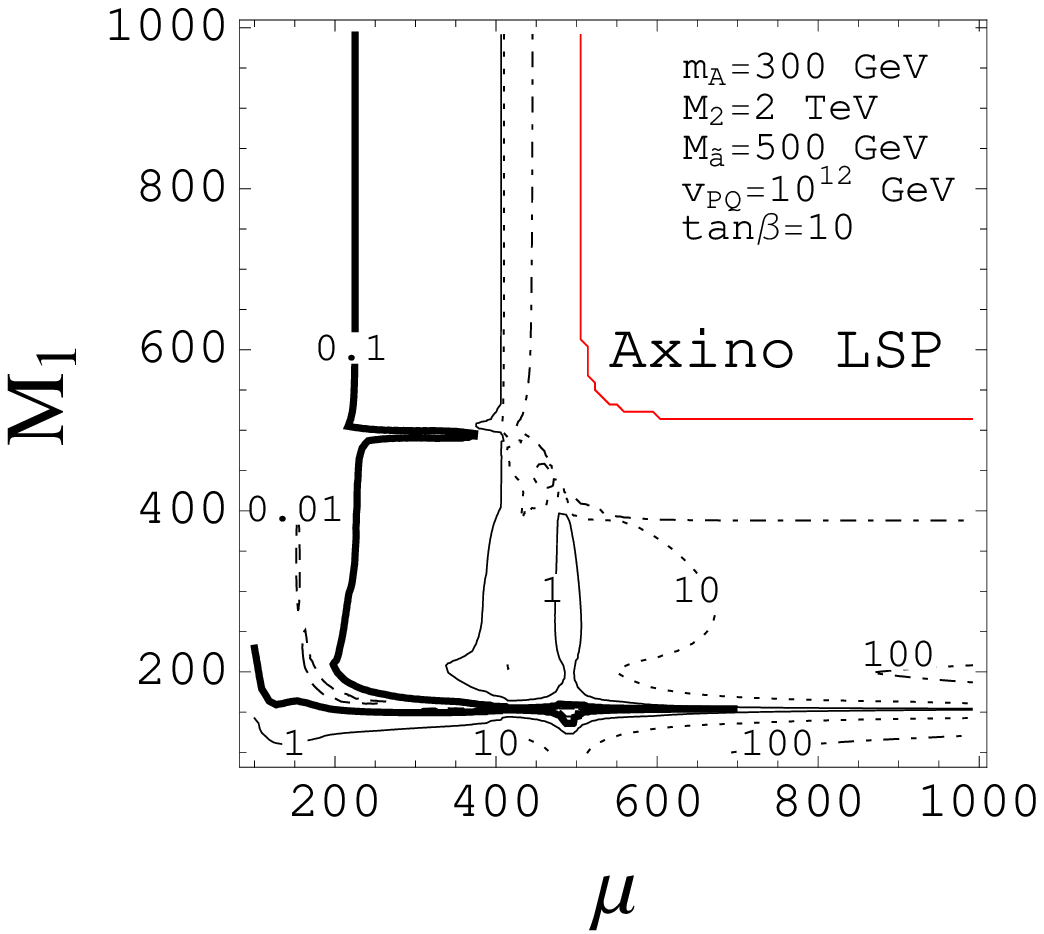}\quad
\includegraphics[width=8cm]{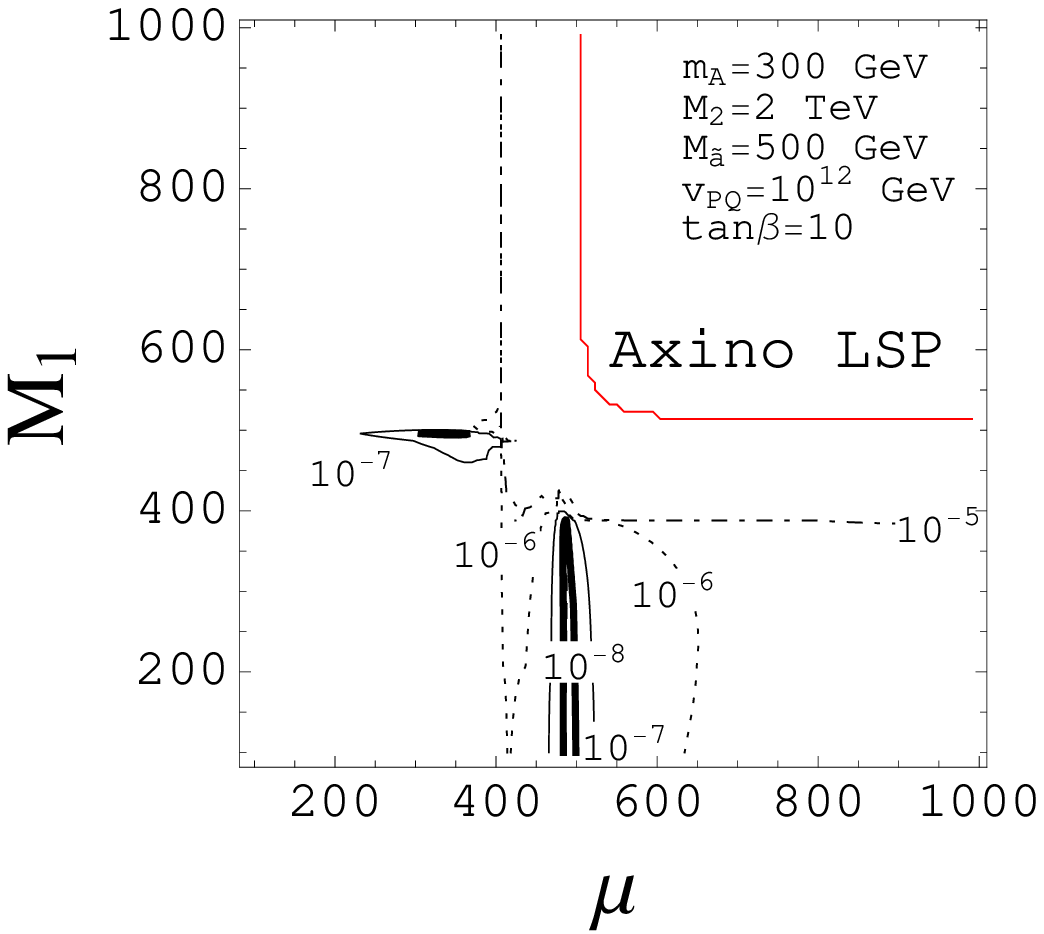}
\caption{Same as the previous figure with $v_{PQ}=10^{12}$ GeV.
\label{fig:1e12_M2_2000}}
\end{figure}

Let us now analyze the relic abundance of the neutralino dark matter
in the simplified MSSM parameter space varying the PQ scale $v_{PQ}$.
Only relevant MSSM parameters are $M_1$, $M_2$ and $\mu$ in our calculation.

In Fig.~\ref{fig:Oh2_MSSM_M2_2000}, we show the contours of the usual freeze-out relic abundance of
the neutralino LSP and the direct detection rate in $M_1$-$\mu$ plane, setting $M_2=2$ TeV for simplicity.
Contours are numerically obtained by using SUSY-HIT \cite{Djouadi:2006bz} and micrOMEGAs 2.4 \cite{Belanger:2010gh}.
Note that the WMAP consistent relic density $\Omega_{DM} h^2 \approx 0.1$ \cite{Komatsu:2010fb}
is obtained in the Higgsino-like LSP region ($M_1\approx \mu$) and the resonant annihilation region
with $M_1 \approx m_A/2$ as expected.  One can also see that the direct detection experiment \cite{Aprile:2011hi} excludes the Higgsino-like region and the lower $\mu$ part of the resonance
region.

Let us proceed to consider the effect of the axino late decay in
Figs.~\ref{fig:1e10_M2_2000}-\ref{fig:1e12_M2_2000}
 for the reheat temperature above the electroweak scale.
We take $m_{\tilde{a}}=500$ GeV and $\tan\beta=10$ for all the plots, and
$v_{PQ}=10^{10}$ GeV in Fig.~\ref{fig:1e10_M2_2000},
$v_{PQ}=10^{11}$ GeV in Fig.~\ref{fig:1e11_M2_2000} and
$v_{PQ}=10^{12}$ GeV in Fig.~\ref{fig:1e12_M2_2000}.

As can be seen in the left panel of  Fig.~\ref{fig:1e10_M2_2000}, the Higgsino-like LSP with $\mu\approx 400$ GeV in the upper-left part gives the correct dark matter relic density.
In this part, the Bino-Higgsino mixing becomes negligible,
and the axino decay is determined by the tree-level axino-Higgino-Higgs coupling (\ref{dfsz-coupling}).
This is why the contour is almost parallel to the vertical axis in the large $M_1$ region.
Note that the phase space of the axino 2-body decay to the Higgsino-like LSP and the Higgs becomes very small for $\mu\approx 400$ GeV.
For $\mu \gtrsim 400$ GeV, the phase space of the axino 2-body decay is closed, but the axino can decay to the Higgsino-like LSP through the processes like 3-body decays with almost massless quarks and leptons via the off-shell Higgses or Z-boson, 2-body decay by the axino-photon-neutralino interaction in Eqs.~(\ref{axino_decay:chi_gamma}), {\it etc}. However all these processes are much suppressed by phase space factors or the 1-loop factor. So in this region, the axino lives long enough to decay after the Higgsino-like LSP freeze-out.
Thus, one can understand that
the vertical line with $\mu\approx 400$ GeV with $M_1 \gtrsim 400$ GeV is the parameter region which
provides a correct dark matter density
with an appropriately enhanced annihilation rate for the given axino lifetime $\tau_{\tilde a} \approx 10^{-8}$ s (as shown in the right panel)  according to the reannihilation equation (\ref{re-ann}). On the other hand one can see that the thick contour line for the correct dark matter density in the lower-right part of the left panel is almost same as in
Fig.~\ref{fig:Oh2_MSSM_M2_2000} as the corresponding axino lifetime is short enough for the axino
to decay before the Bino-like LSP freeze-out.

For $v_{PQ}=10^{11}$ GeV, the axino decays later as seen in the right panel of
Fig.~\ref{fig:1e11_M2_2000}. As a consequence, the contour line for the right dark matter
density changes slightly compared to the previous case. More dramatic change can be found for
$v_{PQ}=10^{12}$ GeV as in Fig.~\ref{fig:1e12_M2_2000}.  The Higgsino-like dark matter becomes
significantly lighter ($\mu\sim 250$ GeV) and the Bino-like dark matter needs more fine-tuned
resonance  both of which guarantee more enhanced reannihilation by a factor of $T_f/T_D$.
Note also that the light Higgsino-like dark matter is allowed by the direct detection
experiment only for the region of  $M_1 \gtr 700$ GeV and
the Bino-like dark matter region requiring stronger resonance is more restricted
(compare with the blue lines in Fig.~\ref{fig:Oh2_MSSM_M2_2000}).

\section{Conclusion}

We examined thermal production of the DFSZ axino and its impact on the dark matter property.
The DFSZ axino interactions are governed by two important couplings: the axino-Higgs-Higgsino
and axino-top-stop coupling proportional to $\mu/v_{PQ}$ and $m_t/v_{PQ}$, respectively.
The latter coupling, which arises from the axino-Higgsino mixing,
is effective only after the electroweak symmetry breaking.
All the scattering, decay and inverse decay processes if allowed are shown to produce
the typical DFSZ axino density of $Y_{\tilde a} \sim 10^{-5} (10^{11} \mbox{GeV}/v_{PQ})^2$
when the reheat temperature is larger than the electroweak scale.
In a certain mass parameter region,
a resonant scattering can occur to allow enhanced axino production.
Thus, if the axino is the LSP, its mass has to be very light
to get the right dark matter density.  If the axino LSP mass is at the TeV scale,
the reheat temperature needs to be below around 50 GeV in order not to overclose the Universe.

A heavy DFSZ axino can decay to a neutralino LSP through the axino-Higgs-Higgsino coupling,
which leads to a huge neutralino population. Its impact is studied to identify the
Bino-Higgsino LSP region consistent with the observed dark matter density and
the direct detection limit of dark matter. For lower PQ scale, a new Higgsino-like LSP region
appears to meet an appropriate reannnihilation condition, whereas the usual Bino-like LSP region
with a resonant annihilation through a CP-odd Higgs boson remains unchanged as the axino decays
before the neutralino freeze-out.
For higher PQ scale, the axino decay occurs much later than the neutralino freeze-out which requires
stronger reannihilation.  As a result, the neutralino dark matter region is more depleted
to a lighter Higgsino-like LSP and a more restricted Bino-like LSP.

\medskip

{\bf Acknowledgement}: The authors thank Prof.\ Kiwoon Choi for his suggestions and
comments on the manuscript. KJB is supported by TJ Park Postdoctoral Fellowship of POSCO TJ Park
Foundation. SHI is supported by the KRF Grants funded by the Korean
Government (KRF-2008-314-C00064 and KRF-2007-341-C00010) and the
KOSEF Grant funded by the Korean Government (No.\ 2009-0080844).
KJB and SHI are also supported by the BK21 project by the Korean Government.

\section*{Appendix}

\subsection{Axino-neutralino mixing}

In the supersymmetric DFSZ axion model, we have
axino-neutralino mixing because of vacuum expectation values of Higgs doublets.
Let us write down the axino-neutralino mass matrix in the basis of ($\tilde{a}$, $\widetilde{B}$, $\widetilde{W}^3$, $\widetilde{H}_d^0$, $\widetilde{H}_u^0$):
\begin{equation}
M=
\begin{pmatrix}
m_{\tilde{a}} & 0 & 0 & c_H\mu vs_{\beta}/v_{PQ} & c_H\mu vc_{\beta}/v_{PQ} \\
0 & M_1 & 0 & -c_{\beta}s_Wm_Z & s_{\beta}s_Wm_Z \\
0 & 0 & M_2 & c_{\beta}c_Wm_Z & -s_{\beta}c_Wm_Z \\
c_H\mu vs_{\beta}/v_{PQ} & -c_{\beta}s_Wm_Z & c_{\beta}c_Wm_Z & 0 & -\mu \\
c_H\mu vc_{\beta}/v_{PQ} & s_{\beta}s_Wm_Z & -s_{\beta}c_Wm_Z & -\mu & 0 \\
\end{pmatrix},
\end{equation}
where $s_{\beta}=\sin\beta$, $c_{\beta}=\cos\beta$, $s_W=\sin\theta_W$, $c_W=\cos\theta_W$
and $v= 174$ GeV.
The above matrix can be diagonalized by the method of perturbative diagonalization as in
Ref.~\cite{Bae:2007pa}:
\begin{equation}
M^{\text{diag}}=NMN^T=VM'V^T=VUMU^TV^T
\end{equation}
where
\begin{equation}
U=
\begin{pmatrix}
1 & 0 & 0 & 0 & 0 \\
0 & 1 & 0 & 0 & 0 \\
0 & 0 & 1 & 0 & 0 \\
0 & 0 & 0 & \frac{1}{\sqrt{2}} & -\frac{1}{\sqrt{2}} \\
0 & 0 & 0 & \frac{1}{\sqrt{2}} & \frac{1}{\sqrt{2}} \\
\end{pmatrix}
\end{equation}
and
\begin{equation}
M'=
\begin{pmatrix}
A & B \\ B^T & C
\end{pmatrix}
\end{equation}
with
\begin{equation}
A=
\begin{pmatrix}
m_{\tilde{a}} & 0 & 0 \\
0 & M_1 & 0 \\
0 & 0 & M_2 \\
\end{pmatrix},~
B=
\begin{pmatrix}
c_H\mu v (s_{\beta}-c_{\beta})/\sqrt{2}v_{PQ} & c_H\mu v (s_{\beta}+c_{\beta})/\sqrt{2}v_{PQ} \\
-m_Zs_W(s_{\beta}+c_{\beta})/\sqrt{2} & m_Zs_W(s_{\beta}-c_{\beta})/\sqrt{2} \\
m_Zc_W(s_{\beta}+c_{\beta})/\sqrt{2} & m_Zc_W(s_{\beta}-c_{\beta})/\sqrt{2} \\
\end{pmatrix},~
C=
\begin{pmatrix}
\mu & 0 \\ 0 & -\mu \\
\end{pmatrix} .
\end{equation}
In the leading order we have
\begin{equation}
V_{nm}=\frac{M'_{mn}}{M_{nn}-M_{mm}}
\end{equation}
As far as $|M_1-\mu|,|m_{\tilde{a}}-\mu|\gg v/v_{PQ},\mu/v_{PQ}$, we find
\begin{eqnarray}
V_{03}&=&\frac{M'_{30}}{m_{\tilde{a}}-\mu}=-\frac{c_H\mu v(s_{\beta}-c_{\beta})}{\sqrt{2}v_{PQ}(\mu-m_{\tilde{a}})},\\
V_{04}&=&\frac{M'_{40}}{m_{\tilde{a}}+\mu}=\frac{c_H\mu v(s_{\beta}+c_{\beta})}{\sqrt{2}v_{PQ}(\mu+m_{\tilde{a}})},\\
V_{13}&=&\frac{M'_{31}}{M_1-\mu}=\frac{m_Zs_W(s_{\beta}+c_{\beta})}{\sqrt{2}(\mu-M_1)},\\
V_{14}&=&\frac{M'_{41}}{M_1+\mu}=\frac{m_Zs_W(s_{\beta}+c_{\beta})}{\sqrt{2}(\mu+M_1)}.
\end{eqnarray}
where the zeroth component denotes the axino component.
Hence we have
\begin{eqnarray}
N_{03}&=&\frac{1}{\sqrt{2}}(V_{03}+V_{04})=-\frac{c_H\mu v(m_{\tilde{a}}s_{\beta}-\mu c_{\beta})}{v_{PQ}(\mu^2-m_{\tilde{a}}^2)},\\ \label{axino-higgsino}
N_{04}&=&\frac{1}{\sqrt{2}}(-V_{03}+V_{04})=\frac{c_H\mu v(\mu s_{\beta}-m_{\tilde{a}} c_{\beta})}{v_{PQ}(\mu^2-m_{\tilde{a}}^2)},\\
N_{13}&=&\frac{1}{\sqrt{2}}(V_{13}+V_{14})=\frac{m_Zs_W(\mu s_{\beta}+M_1 c_{\beta})}{\mu^2-M_1^2},\\
N_{14}&=&\frac{1}{\sqrt{2}}(-V_{13}+V_{14})=-\frac{m_Zs_W(M_1s_{\beta}+\mu c_{\beta})}{\mu^2-M_1^2}.
\end{eqnarray}

Having obtained the mixing matrices $N$, $U$ and $V$, one can now get the following axino couplings
following Ref.~\cite{Jungman:1995df}.

{\bf Axino-neutralino-$Z$ boson coupling}:
\begin{equation}
{\cal L}_{Z\tilde{a}\tilde{\chi}^0}=-\frac{g_2}{c_W}O''_{0n}Z_{\mu}\bar{\tilde{a}}\gamma^{\mu}\gamma^5\tilde{\chi}_n^0+\text{h.c.}
\end{equation}
where
\begin{equation}
O''_{0n}=\frac12(-N_{03}N_{n3}+N_{04}N_{n4}).\label{coupling:O_01}
\end{equation}

{\bf Axino-chargino-$W$ boson coupling}:
\begin{equation}
{\cal L}_{W^{\pm}\tilde{a}\tilde{\chi}^{\mp}}=g_2W^-_{\mu}\bar{\tilde{a}}\gamma^{\mu}\big[O^L_{0n}P_L+O^R_{0n}P_R\big]\tilde{\chi}^+_n+\text{h.c.},
\end{equation}
where
\begin{eqnarray}
O^L_{0n}&=&-\frac{1}{\sqrt{2}}N_{04}V_{n2}+N_{02}V_{n1},\\
O^R_{0n}&=&\frac{1}{\sqrt{2}}N_{03}U_{n2}+N_{02}U_{n1}.
\end{eqnarray}

{\bf Axino-neutralino-neutral Higgs coupling}:

\begin{eqnarray}
{\cal L}_{h\tilde{a}\tilde{\chi}^0}&=&g_2T_{h0n} h\bar{\tilde{a}}\tilde{\chi}^0_n,\\
{\cal L}_{H\tilde{a}\tilde{\chi}^0}&=&g_2T_{H0n}H\bar{\tilde{a}}\tilde{\chi}^0_n,\\
{\cal L}_{A\tilde{a}\tilde{\chi}^0}&=&ig_2T_{A0n}A\bar{\tilde{a}}\gamma^5\tilde{\chi}^0_n,
\end{eqnarray}
where
\begin{eqnarray}
T_{h0n}&=&\sin\alpha Q''_{0n}+\cos\alpha S''_{0n},\label{coupling:T_h0n}\\
T_{H0n}&=&-\cos\alpha Q''_{0n}+\sin\alpha S''_{0n},\label{coupling:T_H0n}\\
T_{A0n}&=&-\sin\beta Q''_{0n}+\cos\beta S''_{0n},\label{coupling:T_A0n}\\
Q''_{0n}&=&\frac12N_{03}(N_{n2}-\tan\theta_W N_{n1})+\frac{c_H\mu}{\sqrt{2}g_2v_{PQ}}N_{00}N_{n4}+(0\leftrightarrow n),\label{eq:Q_01}\\
S''_{0n}&=&\frac12N_{04}(N_{n2}-\tan\theta_W N_{n1})-\frac{c_H\mu}{\sqrt{2}g_2v_{PQ}}N_{00}N_{n3}+(0\leftrightarrow n).\label{eq:S_01}
\end{eqnarray}

{\bf Axino-chargino-charged Higgs coupling}:
\begin{equation}
{\cal L}_{H^{\pm}\tilde{a}\tilde{\chi}^{\mp}}=-g_2H^-\bar{\tilde{a}}\big[Q'^L_{0n}P_L+Q'^R_{0n}P_R\big]\tilde{\chi}^++\text{h.c.},
\end{equation}
where
\begin{eqnarray}
Q'^L_{0n}&=&\cos\beta\big[N_{04}V_{n1}+\frac{1}{\sqrt{2}}(N_{02}+\tan\theta_WN_{01})V_{n2}\big]+\sin\beta\frac{c_H\mu}{g_2v_{PQ}}N_{00}V_{n2}\\
Q'^R_{0n}&=&\sin\beta\big[N_{03}U_{n1}+\frac{1}{\sqrt{2}}(N_{02}+\tan\theta_WN_{01})U_{n2}\big]+\cos\beta\frac{c_H\mu}{g_2v_{PQ}}N_{00}U_{n2}
\end{eqnarray}

\subsection{Axino decay rate}
We first consider the case of $m_{\tilde{a}}<\mu$.
Among the 4 possible decay modes, (\ref{axino_decay:chi_gamma}) is one-loop suppressed and (\ref{axino_decay:3body}) is suppressed by 3 body phase space.
On the other hand, (\ref{axino_decay:H_chi}) and (\ref{axino_decay:Z_chi}) are tree-level processes which arise from the axino-Higgsino and Higgsino-Bino mixing. Being 2-body decays they are the dominant processes.
The decay rates are given by
\begin{eqnarray}
\Gamma(\tilde{a}\to Z+\tilde{\chi})&=&\frac{m_{\tilde{a}}\lambda^{1/2}(1,\kappa^2_Z,\kappa^2_{\chi})}{16\pi }\frac{g_2^2}{c_W^2}|O''_{01}|^2\frac{\big\{(1+\kappa_{\chi})^2-\kappa^2_Z\big\}\big\{(1-\kappa_{\chi})^2+2\kappa^2_Z\big\}}{\kappa^2_Z} \nonumber\\
&\approx&\frac{g_2^2t_W^2}{64\pi}\frac{v^2m_{\tilde{a}}^5}{v_{PQ}^2\mu^4}\label{Gamma:aZBt}\\
\Gamma(\tilde{a}\to h+\tilde{\chi})&=&\frac{m_{\tilde{a}}\lambda^{1/2}(1,\kappa_{h}^2,\kappa_{\chi}^2)}{16\pi}g_2^2|T_{h01}|^2\big\{(1+\kappa_{\chi})^2-\kappa_h^2\big\}\nonumber\\
&\approx&\frac{g_2^2t_W^2}{16\pi}\frac{v^2m_{\tilde{a}}}{v_{PQ}^2}\label{Gamma:ahBt}\\
\Gamma(\tilde{a}\to H+\tilde{\chi})&=&\frac{m_{\tilde{a}}\lambda^{1/2}(1,\kappa_{h}^2,\kappa_{\chi}^2)}{16\pi}g_2^2|T_{H01}|^2\big\{(1+\kappa_{\chi})^2-\kappa_H^2\big\}\nonumber\\
&\approx&\frac{g_2^2t_W^2}{64\pi}\frac{v^2m_{\tilde{a}}^3}{v_{PQ}^2\mu^2}\label{Gamma:aHBt}\\
\Gamma(\tilde{a}\to A+\tilde{\chi})&=&\frac{m_{\tilde{a}}\lambda^{1/2}(1,\kappa_{h}^2,\kappa_{\chi}^2)}{16\pi}g_2^2|T_{A01}|^2\big\{(1-\kappa_{\chi})^2-\kappa_A^2\big\}\nonumber\\
&\approx&\frac{g_2^2t_W^2}{64\pi}\frac{v^2m_{\tilde{a}}^3}{v_{PQ}^2\mu^2}\label{Gamma:aABt}
\end{eqnarray}
where $g_2$ is $SU(2)$ coupling constant, $c_W=\cos\theta_W$, $t_W=\tan\theta_W$, $\theta_W$ is the weak mixing angle, $\lambda^{1/2}(1,a,b)=[(1-a-b)^2-4ab]^{1/2}$,
$\kappa_Z=m_Z/m_{\tilde{a}}$, $\kappa_{\chi}=M_{\tilde{\chi}}/m_{\tilde{a}}$, $\kappa_h=m_h/m_{\tilde{a}}$, $\kappa_H=m_H/m_{\tilde{a}}$, and $\kappa_A=m_A/m_{\tilde{a}}$.
$O''^L_{01}$, $T_{h01}$, $T_{H01}$ and $T_{A01}$ are given in (\ref{coupling:T_h0n}), (\ref{coupling:T_H0n}) and (\ref{coupling:T_A0n}), respectively.
The approximation in the second line of each decay rate is valid for $\kappa_i\ll 1$.

For the case of $m_{\tilde{a}}>\mu$, the axino can directly decay to a Higgsino-like neutralino and a chargino so that the decay rates can be even higher than those in the case of $m_{\tilde{a}}<\mu$.
The axino decay rates are
\begin{eqnarray}
\Gamma(\tilde{a}\to h+\tilde{H}^0)&=&\frac{m_{\tilde{a}}\lambda^{1/2}(1,\kappa_{h}^2,\kappa_{\mu}^2)}{16\pi}g_2^2\big[|T_{h03}|^2+|T_{h04}|^2\big]\big\{(1+\kappa_{\mu})^2-\kappa_h^2\big\}  \nonumber     \\
&\approx&\frac{1}{32\pi}\frac{\mu^2m_{\tilde{a}}}{v_{PQ}^2}\label{Gamma:ahHt}\\
\Gamma(\tilde{a}\to H+\tilde{H}^0)&=&\frac{m_{\tilde{a}}\lambda^{1/2}(1,\kappa_{H}^2,\kappa_{\mu}^2)}{16\pi}g_2^2\big[|T_{H03}|^2+|T_{H04}|^2\big]\big\{(1+\kappa_{\mu})^2-\kappa_H^2\big\}  \nonumber           \\
&\approx&\frac{1}{32\pi}\frac{\mu^2m_{\tilde{a}}}{v_{PQ}^2}\label{Gamma:aHHHt}\\
\Gamma(\tilde{a}\to A+\tilde{H}^0)&=& \frac{m_{\tilde{a}}\lambda^{1/2}(1,\kappa_{A}^2,\kappa_{\mu}^2)}{16\pi}g_2^2\big[|T_{A03}|^2+|T_{A04}|^2\big]\big\{(1-\kappa_{\mu})^2-\kappa_A^2\big\}  \nonumber           \\
&\approx&\frac{1}{32\pi}\frac{\mu^2m_{\tilde{a}}}{v_{PQ}^2}  \label{Gamma:aAHt}    \\
\Gamma(\tilde{a}\to Z+\tilde{H}^0)&=&\frac{m_{\tilde{a}}\lambda^{1/2}(1,\kappa^2_Z,\kappa^2_{\mu})}{16\pi }\frac{g_2^2}{c_W^2}\big[|O''_{03}|^2+|O''_{04}|^2\big]\frac{\big\{(1+\kappa_{\mu})^2-\kappa^2_Z\big\}\big\{(1-\kappa_{\mu})^2+2\kappa^2_Z\big\}}{\kappa^2_Z} \nonumber\\
&\approx&\frac{g_2^2}{64\pi c_W^2}\frac{v^2\mu^2}{v_{PQ}^2m_{\tilde{a}}}     \label{Gamma:a-ZHt}\\
\Gamma(\tilde{a}\to W^{\pm}+\tilde{H}^{\mp})&=&\frac{m_{\tilde{a}}\lambda^{1/2}(1,\kappa^2_W,\kappa^2_{\mu})}{16\pi }g_2^2 \nonumber\\
&&\times\biggl[\big\{|O^L_{01}|^2+|O^L_{01}|^2\big\}\frac{\big\{(1-\kappa_{\mu}^2)^2+(1+\kappa_{\mu}^2-2\kappa_W^2)\kappa_W^2\}}{\kappa^2_Z} \nonumber      \\
&&+\big\{O^L_{01}O^R_{01}\big\}12\kappa_{\mu}
\biggr]\nonumber\\
&\approx&\frac{g_2^2}{32\pi}\frac{v^2\mu^2}{v_{PQ}^2m_{\tilde{a}}}   \label{Gamma:aWHt}  \\
\Gamma(\tilde{a}\to H^{\pm}+\tilde{H}^{\mp})&=&\frac{m_{\tilde{a}}\lambda^{1/2}(1,\kappa_{H^{\pm}}^2,\kappa_{\mu}^2)}{16\pi}g_2^2 \nonumber\\
&&\times\biggl[
\big\{|Q'^L_{01}|^2+|Q'^R_{01}|^2\big\}\big\{1+\kappa_{\mu}^2-\kappa_{H^{\pm}}^2\big\}+\big\{Q'^L_{01}Q'^R_{01}\big\}2\kappa_\mu\biggr] \nonumber     \\
&\approx&\frac{1}{16\pi}\frac{\mu^2m_{\tilde{a}}}{v_{PQ}^2}     \label{Gamma:aHpHt}
\end{eqnarray}
where $\kappa_W=m_W/m_{\tilde{a}}$, $\kappa_{H^{\pm}}=m_{H^{\pm}}/m_{\tilde{a}}$ and $\kappa_{\mu}=\mu/m_{\tilde{a}}$.

\end{document}